\documentclass[preprint,aps,showpacs,nofootinbib,preprintnumbers,amsmath,amssymb]{revtex4-1}
\usepackage{epsfig}
\usepackage{bm}
\usepackage[usenames ,dvipsnames]{xcolor}
\usepackage{slashed}
\usepackage{graphicx,color}

\usepackage{multirow}
\begin{document}
	\title{  Charmed Baryon Weak Decays with Decuplet Baryon and  SU(3) Flavor Symmetry}
	
	\author{Chao-Qiang Geng$^{1,2,3}$, Chia-Wei Liu$^{2}$, Tien-Hsueh Tsai$^{2}$ and Yao Yu$^{1}$\footnote{yuyao@cqupt.edu.cn}}
	\affiliation{
		$^{1}$Chongqing University of Posts \& Telecommunications, Chongqing 400065\\
		$^{2}$Department of Physics, National Tsing Hua University, Hsinchu 300\\
		$^{3}$Physics Division, National Center for Theoretical Sciences, Hsinchu 300
	}\date{\today}
\begin{abstract}
We study the  branching ratios and up-down asymmetries in the charmed baryon weak decays of ${\bf B}_c\to {\bf B}_DM$  with ${\bf B}_{c(D)}$ anti-triplet charmed (decuplet) baryon and $M$ pseudo-scalar meson states based on the flavor symmetry of  $SU(3)_F$. We propose equal and physical-mass schemes for the hadronic states to  deal with the large variations of the  decuplet baryon momenta in the decays in order to fit with the current experimental data. We find that our fitting results of ${\cal B}({\bf B}_c\to {\bf B}_DM) $ are consistent with the current experimental data in both schemes, while the up-down asymmetries in all decays are found to be sizable,  consistent with the current  experimental data, but different from zero predicted in the literature.
We also examine the processes of $\Xi_c^0 \to \Sigma^{\prime 0}K_S/K_L$ and derive  the  asymmetry between the $K_L/K_S$ modes being a constant.
 \end{abstract}

\maketitle

\section{Introduction}

There have been many interesting progresses in the study of charmed baryon weak decays due to
 the recent measurement of the absolute branching fraction for the golden mode $\Lambda_c^+ \to p K^-\pi^+$
by the Belle Collaboration~\cite{Yang:2015ytm} with the new world average of
  ${\cal B}(\Lambda_c^+ \to p K^-\pi^+)=(6.23\pm 0.33)\%$~\cite{pdg} as well as other  $\Lambda_c$ measurements 
  by the BESIII Collaboration~\cite{Ablikim:2015flg,Ablikim:2015prg,Ablikim:2016tze,Ablikim:2016mcr,Ablikim:2017ors,
Ablikim:2016vqd,Ablikim:2017iqd,Ablikim:2018jfs,Ablikim:2018bir,etaAb,SiAbsoluteBr}.
In addition, the absolute branching ratio of ${\cal B}(\Xi_c^0 \to \Xi^-\pi^+)=(1.80\pm0.52)\%$
 is given  by the BELLE collaboration for the first time~\cite{XiAbsoluteBr}.
Theoretically, the charmed baryon decays are dominated  by the nonfactorizable effects, such as 
 those associated with the non-vanishing measured branching ratios for the Cabibbo allowed decays of 
 $\Lambda_c^+ \to \Sigma^0\pi^+$ and $\Lambda_c^+ \to \Sigma^+\pi^0$~\cite{pdg},
 which do not receive any factorizable contributions.
 Many efforts have been made to understand  the nonfactorizable effects  with  different dynamical QCD models~\cite{Heav,Xu92,xialpha1,xialpha2,Cheng:allow,Cheng:general,Uppal:1994pt,Verma98,pole}
 as well as  the use of  the flavor symmetry of 
$SU(3)_F$~\cite{Sharma:1996sc,Savage:1989qr,Savage:1991wu,Lu:2016ogy,first,zero,second,third,fourth,fifth,Wang:2017gxe,sixth,Zhao:2018mov,Hsiao:2019yur,Geng:2019bfz},
which is believed to be the most reliable and simple way to examine the charmed baryon processes.
In particular,  it has been recently demonstrated  that the results for the charmed baryon decays based on the  $SU(3)_F$ 
approach~\cite{Lu:2016ogy,first,zero,second,third,fourth,fifth,Wang:2017gxe,sixth,Zhao:2018mov,Hsiao:2019yur,Geng:2019bfz} are consistent with the experimental data.

 However, most of the recent experimental and theoretical activities  
 have been concentrated on the charmed baryon  decays with the octet baryon in the final states, whereas
 there has been a little  studies for the decuplet modes. 
 Note that most of the charmed baryon experiments with the decuplet baryon were all done before the millennium.
   In this work, we will examine the two-body weak decays of ${\bf B}_c\to {\bf B}_DM$,
   where ${\bf B}_{c(D)}$ and $M$ represent
the anti-triplet charmed (decuplet) baryon and  octet pseudo-scalar meson states
 based on  $SU(3)_F$. 
   There are two important features  for the decays of  ${\bf B}_c\to {\bf B}_DM$.
    The first one is that all factorizable amplitudes vanish, 
    resulting in theoretically clean predictions for the non-factorizable contributions of the decays.
    The other one is that the decays involve only a few $SU(3)_F$ parameters, which 
    are able to be determined  by the current experimental data.

On the other hand,  the up-down asymmetries of $\alpha$ in  $\Lambda_c^+ \to \Xi^{0} K^+$ and $\Lambda_c^+\to  \Xi^{\prime 0} K^+$  have also  been given 
recently by  
the BESIII Collaboration with the results of
 $\alpha(\Lambda_c^+ \to \Xi^{0} K^+, \Xi^{\prime0} K^+)=(0.77\pm0.78, -1.00\pm0.34)$~\cite{Ablikim:2018bir}, respectively,
 where $\Xi^{\prime 0}$ belongs to the decuplet baryon state with spin-3/2. 
 Although the former experimental result is still consistent with zero, the later one is clearly sizable. This non-vanishing large asymmetry 
 is different from the  prediction in the most theoretical calculations in the 
literature~\cite{Heav,Xu92,xialpha1,xialpha2,Cheng:allow,Cheng:general,Uppal:1994pt,Sharma:1996sc,Verma98,pole}. 
Recently, based on the flavor symmetry of $SU(3)_F$, we show that 
$\alpha(\Lambda_c^+ \to \Xi^{0} K^+)=(0.94^{+0.06}_{-0.11})$~\cite{sixth}, which is consistent with the data, but much larger than  zero.
In this work, we will particular check  the up-down asymmetry in $\Lambda_c^+\to  \Xi^{\prime 0} K^+$ to see if it agrees with the experimental non-zero
 result in the $SU(3)_F$ approach.

This paper is organized as the follow. In Sec.~II, we present the formalism.
 We show  how the decay amplitudes are related based on $SU(3)_F$. In Sec.~III, we provide the numerical results of the decay
branching ratios and up-down asymmetries in ${\bf B}_c\to {\bf B}_DM$.
We conclude our study in Sec.~IV.

\section{Formalism}
In order to investigate the two-body decays of the anti-triplet charmed baryon (${\bf B}_c$) to decuplet baryon  (${\bf B}_D$) and octet  pseudoscalar
meson ($M$) states, we write their representations under  the flavor symmetry of $SU(3)_F$ as
\begin{eqnarray}\label{B_10}
&&{\bf B}_{c} = (\Xi_c^0,-\Xi_c^+,\Lambda_c^+)\,,
\nonumber\\
&&{\bf B}_D=\frac{1}{\sqrt{3}}
\left(\begin{array}{ccc}
\left(\begin{array}{ccc}
\sqrt{3}\Delta^{++}&\Delta^+ & \Sigma^{\prime +}\\
\Delta^+ &\Delta^0 & \frac{\Sigma^{\prime 0}}{\sqrt{2}}\\
\Sigma^{\prime +}& \frac{\Sigma^{\prime 0}}{\sqrt{2}}& \Xi^{\prime 0}
\end{array}\right),
\left(\begin{array}{ccc}
\Delta^+ &\Delta^0 & \frac{\Sigma^{\prime 0}}{\sqrt{2}}\\
\Delta^0 &\sqrt{3}\Delta^-& \Sigma^{\prime -}\\
\frac{\Sigma^{\prime 0}}{\sqrt{2}}&\Sigma^{\prime -}&\Xi^{\prime -}
\end{array}\right),
\left(\begin{array}{ccc}
\Sigma^{\prime +}& \frac{\Sigma^{\prime 0}}{\sqrt{2}}&\Xi^{\prime 0}\\
\frac{\Sigma^{\prime 0}}{\sqrt{2}}&\Sigma^{\prime -}&\Xi^{\prime -}\\
\Xi^{\prime 0}&\Xi^{\prime -}&\sqrt{3}\Omega^-
\end{array}\right)
\end{array}\right)\,,
\nonumber\\
&&M=\left(\begin{array}{ccc}
\frac{1}{\sqrt{2}}\pi^0+\frac{1}{\sqrt{6}} \eta  & \pi^+ & K^+\\
\pi^- &-\frac{1}{\sqrt{2}}\pi^0+\frac{1}{\sqrt{6}}\eta &  K^0\\
K^- & \bar K^0&-\frac{2}{\sqrt{6}}\eta
\end{array}\right)\,.
\end{eqnarray}
Here, we have assumed that the physical meson $\eta$ is solely made of the octet state~\cite{pdg}. 
The effective Hamiltonian associated with   $c\to u \bar{d}s$, $c\to u\bar{q}q$ ($q=d,s$) and $c\to u \bar{s}d$   transitions  
is given by~\cite{first}
\begin{eqnarray}\label{Heff}
{\cal H}_{eff}&=&\sum_{i=+,-}\frac{G_F}{\sqrt 2}c_i
\left(V_{cs}^*V_{ud}O^{ds}_i+V_{cd}^*V_{ud} O^{qq}_i+V_{cd}^*V_{us}O^{sd}_i\right),
\nonumber\\
O_\pm^{q_1q_2}&=&{1\over 2}\left[(\bar u q_1)_{V-A}(\bar{q}_2c)_{V-A}\pm (\bar{q}_2 q_1)_{V-A}(\bar u c)_{V-A}\right]\,,
\nonumber\\
O_\pm^{qq}&\equiv & O_\pm^{dd}-O_\pm^{ss}\,,
\end{eqnarray}
where $(|V_{cs}^*V_{ud}|,|V_{cd}^*V_{ud}|,|V_{cd}^*V_{us}|)\simeq (1,s_c,s_c^2)$ with $s_c\equiv \sin\theta_c\approx 0.225$~\cite{pdg},  
$c_i$ (i=+,-) correspond to the Wilson coefficients, $G_F$ is the Fermi constant, 
and $O_\pm^{q_2q_1}$ with  $(\bar q_1 q_2)\equiv\bar q_1\gamma_\mu(1-\gamma_5)q_2$
represent the four-quark operators.

As  $O_\pm$ belong to $\overline{15}$ and $6$ representations under  $SU(3)_F$, respectively, 
which are symmetric and antisymmetric in flavor and color indices,
we  can decompose the effective Hamiltonian in the tensor forms of $H(\overline{15})$ and  $H(6)$,  given by
\begin{eqnarray}
H(\overline{15})^{ij}_k&=&
\left(\begin{array}{ccc}
\left(\begin{array}{ccc}
0&0&0\\
0&0&0\\
0&0&0
\end{array}\right),
\left(\begin{array}{ccc}
0&s_c&1\\
s_c&0&0\\
1&0&0
\end{array}\right),
\left(\begin{array}{ccc}
0&-s_c^2&-s_c\\
-s_c^2&0&0\\
-s_c&0&0
\end{array}\right)
\end{array}\right)\,,
\nonumber\\
H(6)_{ij}&=&\left(
\begin{array}{ccc}
0& 0 & 0\\
0 & 2 & -2s_c\\
0 & -2s_c& 2s_c^2
\end{array}\right)\,,
\end{eqnarray}
respectively, where we have used the convention of  $V_{cd}=-V_{us}=s_c.$\footnote{Note that there is a sign difference between our convention and 
the one in Ref.~\cite{pdg}, which will not affect our numerical results.}

The most significant feature for  ${\bf B}_c\to {\bf B}_DM$
is that the decay amplitude is essentially non-factorizable due to the vanishing matrix element of the baryonic transition, $i.e.$ $\langle {\bf B}_D|\bar{q}\gamma_\mu(1-\gamma_5)c|{\bf B}_c\rangle=0$. The reason is that the light quark pair in the anti-triplet charmed baryon state  is anti-symmetric, 
whereas that in the decuplet one is  totally symmetric. 
As a result, we can safely neglect  $H(\overline{15})$, which only contributes to the factorizable processes~\cite{quarkscheme,pole,Pati:1970fg,Kohara:1991ug,Korner}. In general, the decay amplitude of ${\bf B}_c\to {\bf B}_DM$ is given by
\begin{equation}\label{MatrixElement}
{\bf M}=\langle {\bf B}_D M |H(6)|{\bf B}_c\rangle=iq_\mu \overline{w}^\mu_{{\bf B}_D}(P-D\gamma_5)u_{{\bf B}_c}\,, 
\end{equation}
where $q_{\mu}$ is the four-momentum of the meson, $w^\mu_{{\bf B}_D}$ is the Rarita-Schwinger spinor vector for the spin-$3/2$ particle
of ${\bf B}_D$, $P(D)$ corresponds to the $P(D)$-wave amplitude, and $u_{{\bf B}_c}$ is the spin-$1/2$ Dirac spinor of ${\bf B}_c$. 
By assuming CP invariance, $P$ and $D$ can be taken to be real. Under $SU(3)_F$, the amplitudes associated with $P$ and $D$ 
are related by
\begin{eqnarray}\label{SU(3)relation}
&&P_{({\bf B}_c\to {\bf B}_DM)} =P_0 f_{{\bf B}_c{\bf B}_DM} \,,~~~~~~~D_{({\bf B}_c\to {\bf B}_DM)} =D_0  f_{{\bf B}_c{\bf B}_DM}  \,,
\end{eqnarray}
respectively,
where $P_0(D_0)$ is the real parameter to be determined and 
$f_{{\bf B}_c{\bf B}_DM} $ is  the $SU(3)_F$ overlapping factor, defined by
\begin{equation}
\label{Ofactor}
 f_{{\bf B}_c{\bf B}_DM}  = ({\bf B}_D)_{ijk}({\bf B}_c)_lH(6)_{om}M^i_q\epsilon^{jlo}\epsilon^{kmq}\,.
\end{equation}
The value of $f_{{\bf B}_c{\bf B}_DM} $ in Eq.~(\ref{Ofactor}) depends on the specific mode in ${\bf B}_c\to {\bf B}_DM$, 
for example, 
\begin{equation}
f_{\Lambda_c^+ \Delta^{++}K^-}= (\Delta^{++})_{111}(\Lambda_c^+)_3 H(6)_{22}(K^-)^1_3\epsilon^{132}\epsilon^{123}=-2\,.
\end{equation}
We will list the values of $f_{{\bf B}_c{\bf B}_DM} $  in the next section.
We note that the factors of $f_{{\bf B}_c{\bf B}_DM} $ with $M$ being a singlet under $SU(3)_F$ vanish so that
the corresponding decays 
with a physical meson $\eta^\prime$ are suppressed.
 The reason is 
 that $\overline{{\bf 3}}\otimes \overline{{\bf 6}} \otimes \overline{{\bf 10}} \otimes {\bf 1}$ cannot form a singlet 
 to be invariant under $SU(3)_F$, where $\overline{{\bf 3}}$, $\overline{{\bf 6}} $, $\overline{{\bf 10}}$ and ${\bf 1} $  are the $SU(3)_F$  representations
  for the anti-triplet charmed baryon, anti-symmetric part of  the effective Hamiltonian, decuplet baryon and  singlet meson states, respectively.
In practice, since $P$ and $D$ share the same overlapping factor, one can alternatively combine these 
two real parameters  into one complex parameter for 
convenience~\cite{Lu:2016ogy,first,zero,second,third,fourth,fifth,Wang:2017gxe,sixth,Zhao:2018mov,Hsiao:2019yur,Geng:2019bfz}.

Consequently, 
the  decay width ($\Gamma$) for ${\bf B}_c\to {\bf B}_DM$ is  given by
\begin{eqnarray}\label{defBr}
\Gamma({\bf B}_c\to {\bf B}_DM)&=&\frac{|\vec{q}|}{8\pi m_{{\bf B}_c}^2}|\langle \overline{{\bf M}^2}\rangle|
=\zeta \left(1+\xi^2r^2\right)P_0^2f^2_{{\bf B}_c{\bf B}_DM}\,,
\end{eqnarray}
while the up-down asymmetry ($\alpha$) has the form
\begin{equation}\label{defasym}
\alpha({\bf B}_c\to {\bf B}_DM)=\frac{2\xi\text{Re} (PD^*)}{|P|^2+\xi^2|D|^2}=\frac{2\xi r}{1+\xi^2 r^2}\,,
\end{equation}
where $|\vec{q}|$ represents the absolute value of the three-momentum of the octet pesudoscalar meson $M$ or the decuplet baryon ${\bf B}_D$ in the CM frame, 
$m_{{\bf B}_c}$ is the mass of the charmed baryon ${\bf B}_c$, 
$|\langle \overline{{\bf M}^2}\rangle|$ stands for the spin average squared amplitude,
$\zeta=(E_{{\bf B}_D}+m_{{\bf B}_D})|\vec{q}|^3m_{{\bf B}_c}/(6\pi m_{{\bf B}_D}^2)$
with $E_{{\bf B}_D}(m_{{\bf B}_D})$ representing energy (mass) of  ${\bf B}_D$,
$\xi=|\vec{q}|/(E_{{\bf B}_D}+m_{{\bf B}_D})$, 
and $r=D_0/P_0$.   

Under the exact flavor symmetry of $SU(3)_F$, one can simply impose the equal-mass ($em$) conditions, given by
\begin{eqnarray}
&&m_{{\bf B}_c}\equiv m_{\Lambda_{c}^+} = m_{\Xi_c^0}= m_{\Xi_c^+}\,,\nonumber\\
&&m_{{\bf B}_D}\equiv m_{\Delta^{++}}=m_{\Delta^+}=m_{\Delta^0}=m_{\Delta^-}=m_{\Sigma^{\prime 0}}=m_{\Sigma^{\prime +}}=m_{\Sigma^{\prime -}}=m_{\Xi^{\prime 0}}=m_{\Xi^{\prime -}}=m_{\Omega^{-}}\,,\nonumber\\
&&m_{M}\equiv m_{\pi^{+}}=m_{\pi^{0}}=m_{\pi^{-}}=m_{K^{+}}=m_{K^{-}}=m_{K^0}=m_{\bar{K}^0}=m_\eta\,,
\end{eqnarray}
leading to that both $\xi$ and $\zeta$ are the same for all decays of ${\bf B}_c\to {\bf B}_DM$.
As a result, 
$\Gamma({\bf B}_c\to {\bf B}_DM)/f^2_{{\bf B}_c{\bf B}_DM}$ and $\alpha({\bf B}_c\to {\bf B}_DM)$ are the same
for  all modes of ${\bf B}_c\to {\bf B}_DM$ when the $em$ conditions are chosen.
This $em$ scheme has been widely used in the charmed baryon decays with the octet baryon in the final states based on $SU(3)_F$ as shown
in Refs.~\cite{Lu:2016ogy,first,zero,second,third,fourth,fifth,Wang:2017gxe,sixth,Zhao:2018mov,Hsiao:2019yur,Geng:2019bfz}.
However, it is clear that both parameters of $\xi$ and $\zeta$ for the  decuplet modes are quite different 
since the three-momentum $\vec{q}$ varies largely in different decays around 0.4-0.8 GeV
 when the physical masses of the baryon and meson states are taken, referred to the physical-mass $(pm)$ scheme.

\section{Numerical Results}

In the  $em$ scheme,  from Eq.~(\ref{defasym}) we see that there is only one combined parameter of $\bar{r}\equiv \xi r$  for 
$\alpha({\bf B}_c\to {\bf B}_DM)$.
By using the experimental data of  $\alpha(\Lambda_c^+\to\Xi^{\prime 0}K^+)=-1.00\pm0.34$ in Ref.~\cite{Ablikim:2018bir}, 
we expect that the up-down asymmetry in every decay mode of ${\bf B}_c\to {\bf B}_DM$ should have the same value as
\begin{equation}\label{alphaem}
	\alpha_{em}({\bf B}_c \to {\bf B}_D M)= -1.00^{+ 0.34}_{-0}\,,
\end{equation}
where  the lower uncertainty of ``0'' reflects that the physical value of $\alpha$ cannot be less than $-1$.
From  Eqs.~(\ref{defasym}) and (\ref{alphaem}), we obtain 
\begin{equation}\label{rem}
	\bar{r}_{em} = -1.00^{+ 0.6}_{-1.6}\,.
\end{equation}
On the other hand, the decay branching ratios in Eq.~(\ref{defBr}) also depend on  one unknown parameter, defined by
\begin{equation}\label{barP0}
\overline{P}_0\equiv \sqrt{\zeta \left(1+\xi^2r^2\right)}P_0\,,
\end{equation}
which can be determined by only  one experimental data point.
However, there are four experimental branching ratios as shown in Table~\ref{table1}. 
\begin{table}
\caption{Our results of  the up-down asymmetries of $\alpha_{pm}$ and branching ratios of ${\cal B}_{pm}$ and ${\cal B}_{em}$ for
the  Cabibbo allowed modes of ${\bf B}_c \to {\bf B}_D M$ based on $SU(3)_F$ along with the 
experimental data~\cite{pdg,Ablikim:2018bir,SiAbsoluteBr,XiAbsoluteBr}
as well as the theoretical calculations in the literature~\cite{pole,Heav,Sharma:1996sc}.}	
	\begin{center}
		\begin{tabular}[t]{|l|c|ccc|ccc|c|}
\hline
channel &$f_{{\bf B}_c{\bf B}_DM} $ & $\alpha_{pm} $ &$10^3{\cal B}_{pm}$& $10^3{\cal B}_{em}$&$10^3{\cal B}$&$10^3{\cal B}$ &$10^3{\cal B} $&$10^3{\cal B}_{ex}(\alpha_{ex})$\\
		      &&\multicolumn{3}{c|}{our result}& \cite{pole}& \cite{Heav}&\cite{Sharma:1996sc}&data\\
\hline
	$ \Lambda_{c}^{+}  \to  \Delta^{++} K^{-} $&$ -2 $&$   -0.86 ^{+ 0.44 }_{- 0.14 }$&$ 15.3 \pm 2.4 $&$ 12.4 \pm 1.0 $&$9.5$&27.0&$7.0\pm4.0$&$10.7\pm2.5$~\cite{pdg}\\
	$ \Lambda_{c}^{+}  \to  \Delta^{+} \bar{K}^{0} $&$ - \frac{2\sqrt{3}}{3} $&$ -0.86 ^{+ 0.44 }_{- 0.14 } $&$ 5.1 \pm 0.8 $&$ 4.1 \pm 0.3 $&$3.1$&9.0&$2.3\pm1.3$&\\
	$ \Lambda_{c}^{+}  \to  \Sigma'^{+} \pi^{0} $&$ \frac{\sqrt{6}}{3} $&$-0.91 ^{+ 0.45 }_{- 0.10 }$&$ 2.2 \pm 0.4 $&$ 2.1 \pm 0.2 $&$2.1$&5.0&$4.6\pm1.8$&\\
	$ \Lambda_{c}^{+}  \to  \Sigma'^{+} \eta $&$\sqrt{2} $&$  -0.97 ^{+ 0.43 }_{- 0.03 }$&$ 3.1 \pm 0.6 $&$ 6.2 \pm 0.5 $&-&10.4&$2.1\pm1.1$&$9.1\pm 2.0$~\cite{SiAbsoluteBr}\\
	$ \Lambda_{c}^{+}  \to  \Sigma'^{0} \pi^{+} $&$ \frac{\sqrt{6}}{3} $&$ -0.90 ^{+ 0.45 }_{- 0.10 }$&$ 2.2 \pm 0.4 $&$ 2.1 \pm 0.2 $&$2.1$&5.0&$4.6\pm1.8$&\\
$ \Lambda_{c}^{+}  \to  \Xi'^{0} K^{+} $&$ \frac{2 \sqrt{3}}{3} $&$ -1.00 ^{+ 0.34 }_{- 0.00 } $&$ 1.0 \pm 0.2 $&$ 4.1 \pm 0.3 $&$0.7$&5.0&$2.3\pm0.9$&$5.02\pm1.04$~\cite{Ablikim:2018bir}\\
			& && & && &&$(-1.00\pm0.34)$~\cite{Ablikim:2018bir}\\
\hline
$ \Xi_{c}^{0}  \to  \Sigma'^{+} K^{-} $&$ \frac{2 \sqrt{3}}{3} $&$  -0.88 ^{+ 0.45 }_{- 0.12 }$&$ 3.1 \pm 0.5 $&$ 2.3 \pm 0.2 $&$1.3$&4.9&$1.3\pm0.7$&\\
$ \Xi_{c}^{0}  \to  \Sigma'^{0} \bar{K}^{0} $&$ \frac{\sqrt{6}}{3} $&$  -0.88 ^{+ 0.45 }_{- 0.12 } $&$ 1.6 \pm 0.2 $&$ 1.2 \pm 0.1 $&$0.7$&2.6&$0.6\pm0.4$&\\
$ \Xi_{c}^{0}  \to  \Xi'^{0} \pi^{0} $&$ -\frac{\sqrt{6}}{3} $&$-0.91 ^{+ 0.45 }_{- 0.09 } $&$ 1.4 \pm 0.2 $&$ 1.2 \pm 0.1 $&$0.9$&2.8&$2.6\pm1.0$&\\
$ \Xi_{c}^{0}  \to  \Xi'^{0} \eta $&$ - \sqrt{2}$&$ -0.97 ^{+ 0.42 }_{- 0.03 }$&$ 2.1 \pm 0.4 $&$ 3.5 \pm 0.3 $&-&0.2&$1.3\pm0.6$&\\
$ \Xi_{c}^{0}  \to  \Xi'^{-} \pi^{+} $&$ - \frac{2 \sqrt{3}}{3} $&$ -0.91 ^{+ 0.45 }_{- 0.09 }$&$ 2.8 \pm 0.5 $&$ 2.3 \pm 0.2 $&$1.9$&5.6&$5.0\pm2.0$&\\
$ \Xi_{c}^{0}  \to  \Omega^{-} K^{+} $&$ -2 $&$  -1.00 ^{+ 0.34 }_{- 0.00 }$&$ 2.3 \pm 0.5 $&$ 7.0 \pm 0.6 $&$1.1$&3.4&$4.5\pm1.8$&$5.4\pm1.6$~\cite{pdg,XiAbsoluteBr}\\
\hline
$\Xi_c^+ \to \Sigma'^+ \bar{K}^0$ &0&-&0&0&0&0&0&$1.0\pm0.5$~\footnote{The data has not been included in the data fitting.}$^\text{b}$~\cite{pdg}\\
$\Xi_c^+ \to \Xi'^0 \pi^+$ &0&-&0&0&0&0&0&$<0.10$\footnote{The experimental decay branching ratios of $\Xi_c^+$ are measured relative to ${\cal B}(\Xi_c^+\to\Xi^-2\pi^+)$.}~\cite{pdg}\\
\hline
\end{tabular}
\label{table1}
\end{center}
\end{table}
To obtain the most plausible value for $\overline{P}_0$
under the current experimental data, we adopt the minimal $\chi^2$ fitting, defined by 
\begin{equation}\label{chi1}
\chi^2_{em}=\sum_{ex} \frac{({\cal B}_{ex}-{\cal B}_{em})^2}{\sigma_{ex}^2}\,,
\end{equation}
where ${\cal B}_{em}$ is the decay branching ratio generated by  $\overline{P}_0$ in the $em$ scheme 
with the experimental measured lifetime in Ref.~\cite{pdg}
and
${\cal B}_{ex}(\sigma_{ex})$ corresponds to the measured branching ratio (uncertainty) in the data. 
By performing $\chi^2$ fit with  the minimal value of $\chi^2_{em}$, we obtain that
\begin{eqnarray}\label{fit1}
(\overline{P}_0)_{em} &=& (8.7\pm 0.5)\times 10^{-3}G_F\text{GeV}^{\frac{5}{2}}\,,
\nonumber\\
\chi^2_{em}/d.o.f. &=& 1.4\,,
\end{eqnarray}
where  $d.o.f.$ represents ``degree of freedom.'' 
The small value of $\chi^2_{em}/d.o.f.$ for the fit in Eq.~(\ref{fit1}) indicates that the $em$ scheme is good to explain the current 
experimental data. Our results for the decay branching ratios of ${\bf B}_c\to {\bf B}_DM$ in the $em$ scheme are summarized 
in Tables~\ref{table1}-\ref{table4}.
In the tables, we also show the $SU(3)_F$ overlapping factors of $f_{{\bf B}_c{\bf B}_DM} $ for the decays of ${\bf B}_c\to {\bf B}_DM$.

We now discuss  in the $pm$ scheme.
From the data of $\alpha_{ex}(\Lambda_c^+\to\Xi^{\prime 0}K^+)$, we find that
\begin{equation}\label{rpm}
	r_{pm} = -6.6^{+ 4.1}_{-10.7}\,.
\end{equation}
With the value  in Eq.~(\ref{rpm}), our predictions for $\alpha_{pm}({\bf B}_c\to {\bf B}_DM)$ are shown in Tables~\ref{table1}-\ref{table4}.
To valuate the decay branching ratios, we have to refit the data, found to be
\begin{eqnarray}\label{fit2}
(P_0,D_0)_{pm}&=&(3.2\pm0.4, -5.1\pm 2.5)10^{-2}G_F \text{GeV}\,, ~~R_{P_0D_0}=0.70\,,
\nonumber\\	
\chi^2_{pm}/d.o.f. &=& 11\,,
\end{eqnarray}
where $R_{P_0D_0}$ stands for the correlation between the two fitted parameters
of $P_0$ and $D_0$ and the data point of $\alpha_{ex}(\Lambda_c^+\to\Xi^{\prime 0}K^+)$ has also been included in the fit.
Our results of ${\cal B}({\bf B}_c\to {\bf B}_DM)$ are listed in Tables~\ref{table1}-\ref{table4}.
We note that unlike the cases in the $em$ scheme,
$\zeta$  and $\xi$  vary from $(0.02  \sim 0.11)~\text{GeV}^3$ and $ (0.2 \sim 0.3)$ for the different modes of
${\bf B}_c \to {\bf B}_D M$ in the $pm$ scheme,
respectively, 
resulting in  the main differences for the two schemes.
In the $pm$ scheme, it is clear that the $SU(3)_F$ flavor symmetry is broken by the mass difference in the kinematic part,
but still kept in the $P(D)$-wave amplitude. 
 In contrast,  $SU(3)_F$  is exact both kinematically and dynamically in  the $em$ scheme.
The larger value of $\chi^2_{pm}$ compared to that of $\chi^2_{em}$ 
 may result from the improper handling of the $P(D)$-wave amplitude.
 The $SU(3)_F$ breaking effect in the amplitude would compensate that from the  kinematic part.
 However, such effect can be considered within the $SU(3)_F$ approach only when more experimental data points are available in the future.

\begin{table}
		\caption{Results for the Cabibbo suppressed decays of	
		${\bf B}_c \to {\bf B}_D M$ with  $SU(3)_F$. }
		\begin{center}
		\begin{tabular}[t]{lccccc}
			\hline
			channel &$s_c^{-1}f_{{\bf B}_c{\bf B}_DM} $&$\alpha_{pm}$ &$10^4{\cal B}_{pm}$&$10^4{\cal B}_{em}$\\
			
			\hline
$ \Lambda_{c}^{+}  \to  \Delta^{++} \pi^{-} $&$ -2 $&$ -0.81 ^{+ 0.43 }_{- 0.19 }$&$ 12.5 \pm 2.0 $&$ 6.6 \pm 0.6 $\\
$ \Lambda_{c}^{+}  \to  \Delta^{+} \pi^{0} $&$ \frac{2 \sqrt{6}}{3} $&$ -0.81 ^{+ 0.43 }_{- 0.19 }$&$ 8.3 \pm 1.3 $&$ 4.4 \pm 0.4 $\\
$ \Lambda_{c}^{+}  \to  \Delta^{0} \pi^{+} $&$ \frac{2 \sqrt{3}}{3} $&$ -0.81 ^{+ 0.43 }_{- 0.19 }$&$ 4.2 \pm 0.7 $&$ 2.2 \pm 0.2 $\\
$ \Lambda_{c}^{+}  \to  \Sigma'^{+} K^{0} $&$ - \frac{2 \sqrt{3}}{3} $&$ -0.95 ^{+ 0.44 }_{- 0.05 }$&$ 1.3 \pm 0.2 $&$ 2.2 \pm 0.2 $\\
$ \Lambda_{c}^{+}  \to  \Sigma'^{0} K^{+} $&$ \frac{\sqrt{6}}{3} $&$ -0.95 ^{+ 0.44 }_{- 0.05 }$&$ 0.7 \pm 0.1 $&$ 1.1 \pm 0.1 $\\
			\hline
$ \Xi_{c}^{0}  \to  \Delta^{+} K^{-} $&$ \frac{2 \sqrt{3}}{3} $&$ -0.79 ^{+ 0.43 }_{- 0.21 }$&$ 3.0 \pm 0.5 $&$ 1.2 \pm 0.1 $\\
$ \Xi_{c}^{0}  \to  \Delta^{0} \bar{K}^{0} $&$ \frac{2 \sqrt{3}}{3} $&$ -0.79 ^{+ 0.43 }_{- 0.21 }$&$ 3.0 \pm 0.5 $&$ 1.2 \pm 0.1 $\\
$ \Xi_{c}^{0}  \to  \Sigma'^{+} \pi^{-} $&$ \frac{2 \sqrt{3}}{3} $&$ -0.84 ^{+ 0.44 }_{- 0.16 }$&$ 2.5 \pm 0.4 $&$ 1.2 \pm 0.1 $\\
$ \Xi_{c}^{0}  \to  \Sigma'^{0} \pi^{0} $&$ - \sqrt{3} $&$ -0.84 ^{+ 0.44 }_{- 0.16 }$&$ 5.6 \pm 0.9 $&$ 2.8 \pm 0.2 $\\
$ \Xi_{c}^{0}  \to  \Sigma'^{0} \eta $&$ -1 $&$ -0.89 ^{+ 0.45 }_{- 0.11 }$&$ 1.1 \pm 0.2 $&$ 0.9 \pm 0.1 $\\
$ \Xi_{c}^{0}  \to  \Sigma'^{-} \pi^{+} $&$ - \frac{4 \sqrt{3}}{3} $&$ -0.84 ^{+ 0.44 }_{- 0.16 }$&$ 9.9 \pm 1.6 $&$ 4.9 \pm 0.4 $\\
$ \Xi_{c}^{0}  \to  \Xi'^{0} K^{0} $&$ \frac{2 \sqrt{3}}{3} $&$ -0.96 ^{+ 0.43 }_{- 0.04 }$&$ 0.9 \pm 0.2 $&$ 1.2 \pm 0.1 $\\
$ \Xi_{c}^{0}  \to  \Xi'^{-} K^{+} $&$ - \frac{4 \sqrt{3}}{3} $&$ -0.96 ^{+ 0.43 }_{- 0.04 }$&$ 3.6 \pm 0.6 $&$ 4.9 \pm 0.4 $\\
\hline
$ \Xi_{c}^{+}  \to  \Delta^{++} K^{-} $&$ 2 $&$ -0.79 ^{+ 0.43 }_{- 0.21 }$&$ 35.0 \pm 5.7 $&$ 14.6 \pm 1.2 $\\
$ \Xi_{c}^{+}  \to  \Delta^{+} \bar{K}^{0} $&$ \frac{2 \sqrt{3}}{3} $&$ -0.79 ^{+ 0.43 }_{- 0.21 }$&$ 11.7 \pm 1.9 $&$ 4.9 \pm 0.4 $\\
$ \Xi_{c}^{+}  \to  \Sigma'^{+} \pi^{0} $&$ - \frac{\sqrt{6}}{3} $&$ -0.84 ^{+ 0.44 }_{- 0.16 }$&$ 4.8 \pm 0.8 $&$ 2.4 \pm 0.2 $\\
$ \Xi_{c}^{+}  \to  \Sigma'^{+} \eta $&$ - \sqrt{2} $&$ -0.89 ^{+ 0.45 }_{- 0.11 }$&$ 8.7 \pm 1.4 $&$ 7.3 \pm 0.6 $\\
$ \Xi_{c}^{+}  \to  \Sigma'^{0} \pi^{+} $&$ - \frac{\sqrt{6}}{3} $&$ -0.84 ^{+ 0.44 }_{- 0.16 }$&$ 4.8 \pm 0.8 $&$ 2.4 \pm 0.2 $\\
$ \Xi_{c}^{+}  \to  \Xi'^{0} K^{+} $&$ - \frac{2 \sqrt{3}}{3} $&$ -0.96 ^{+ 0.43 }_{- 0.04 }$&$ 3.5 \pm 0.6 $&$ 4.9 \pm 0.4 $\\
			\hline
			\label{table2}		
		\end{tabular}
	\end{center}
\end{table}
\begin{table}
		\caption{Results for the Double Cabibbo suppressed decays of	
		${\bf B}_c \to {\bf B}_D M$ with  $SU(3)_F$.}
		\begin{center}
		\label{table3}
		\begin{tabular}[t]{ccccc}
			\hline
			channel &$s_c^{-2}f_{{\bf B}_c{\bf B}_DM} $ &$\alpha_{pm}$&$10^5{\cal B}_{pm}$&$10^5{\cal B}_{em}$\\
			\hline
$ \Xi_{c}^{0}  \to  \Delta^{+} \pi^{-} $&$ \frac{2 \sqrt{3}}{3} $&$ -0.75 ^{+ 0.42 }_{- 0.25 }$&$ 2.2 \pm 0.4 $&$ 0.7 \pm 0.1 $\\
$ \Xi_{c}^{0}  \to  \Delta^{0} \pi^{0} $&$ - \frac{2 \sqrt{6}}{3} $&$ -0.75 ^{+ 0.42 }_{- 0.25 }$&$ 4.3 \pm 0.7 $&$ 1.3 \pm 0.1 $\\
$ \Xi_{c}^{0}  \to  \Delta^{-} \pi^{+} $&$ -2 $&$ -0.75 ^{+ 0.42 }_{- 0.25 }$&$ 6.5 \pm 1.1 $&$ 2.0 \pm 0.2 $\\
$ \Xi_{c}^{0}  \to  \Sigma'^{0} K^{0} $&$ \frac{\sqrt{6}}{3} $&$ -0.88 ^{+ 0.45 }_{- 0.12 }$&$ 0.4 \pm 0.1 $&$ 0.3 \pm 0.0 $\\
$ \Xi_{c}^{0}  \to  \Sigma'^{-} K^{+} $&$ - \frac{2 \sqrt{3}}{3} $&$ -0.88 ^{+ 0.45 }_{- 0.12 }$&$ 0.9 \pm 0.1 $&$ 0.7 \pm 0.1 $\\
\hline
$ \Xi_{c}^{+}  \to  \Delta^{++} \pi^{-} $&$ 2 $&$ -0.76 ^{+ 0.42 }_{- 0.24 }$&$ 25.5 \pm 4.4 $&$ 7.8 \pm 0.7 $\\
$ \Xi_{c}^{+}  \to  \Delta^{+} \pi^{0} $&$ - \frac{2 \sqrt{6}}{3} $&$ -0.76 ^{+ 0.42 }_{- 0.24 }$&$ 17.0 \pm 2.9 $&$ 5.2 \pm 0.4 $\\
$ \Xi_{c}^{+}  \to  \Delta^{0} \pi^{+} $&$ - \frac{2 \sqrt{3}}{3} $&$ -0.76 ^{+ 0.42 }_{- 0.24 }$&$ 8.5 \pm 1.5 $&$ 2.6 \pm 0.2 $\\
$ \Xi_{c}^{+}  \to  \Sigma'^{+} K^{0} $&$ \frac{2 \sqrt{3}}{3} $&$ -0.88 ^{+ 0.45 }_{- 0.12 }$&$ 3.5 \pm 0.6 $&$ 2.6 \pm 0.2 $\\
$ \Xi_{c}^{+}  \to  \Sigma'^{0} K^{+} $&$ - \frac{\sqrt{6}}{3} $&$ -0.88 ^{+ 0.45 }_{- 0.12 }$&$ 1.7 \pm 0.3 $&$ 1.3 \pm 0.1 $\\
			\hline
		\end{tabular}
	\end{center}
\end{table}

In Table~\ref{table1}
for the  Cabibbo allowed modes of ${\bf B}_c \to {\bf B}_D M$ based on $SU(3)_F$,
 we also show the experimental data~\cite{pdg,Ablikim:2018bir,SiAbsoluteBr,XiAbsoluteBr}
as well as the theoretical calculations in the literature~\cite{pole,Heav,Sharma:1996sc}.
In particular,  
Xu and Kamal in Ref.~\cite{pole} consider the baryon pole term as the nonfactorizable amplitude,  
Korner and Kramer in Ref.~\cite{Heav} take account of the heavy quark symmetry and covariant quark model for the baryon wave function, 
and Sharma and Verma in Ref.~\cite{Sharma:1996sc}  study the branching ratios with $SU(3)_F$ based on the old experimental data. 
Note that our result of ${\cal B}_{em}(\Lambda_c^+ \to \Xi^{\prime 0} K^+)=(4.1\pm0.3)\times 10^{-3}$ is smaller than, but still consistent with, 
the current experimental value of $(5.02\pm1.04)\times 10^{-3}$. 
However, it fits well with the previous experimental result of  ${\cal B}(\Lambda_c^+\to \Xi^{\prime 0}K^+)=(4.0 \pm 1.0)\times10^{-3}$ as shown in Table 1 of Ref.~\cite{Ablikim:2018bir}. 
However, our result of  ${\cal B}_{pm}(\Lambda_c^+ \to \Xi^{\prime 0} K^+)=(1.0\pm0.2)\times 10^{-3}$ is inconsistent with the data.
It is interesting to point out that the up-down asymmetries  for all decays are expected to be zero by theoretical studies in  Refs.~\cite{Heav,pole,Sharma:1996sc}
due to the vanishing D-wave amplitudes, which are  different from our nonzero results and inconsistent with
 the current experimental result of  $\alpha_{ex}(\Lambda_c^+ \to \Xi^{\prime 0} K^+)=-1.00\pm0.34$~\cite{Ablikim:2018bir}.
We recommend to measure  $\alpha(\Lambda_c^+ \to \Delta^{++} K^-)$ in the future experiment as this decay channel has the largest
decay branching rate, which will be a clean justification of the $SU(3)_F$ approach.
 In addition,
the authors in Ref.~\cite{Sharma:1996sc} use $SU(3)_F$ without neglecting  $H(\overline{15})$ but  treated the D-wave amplitude being zero. 
Nonetheless, they still arrive the conclusion that $H(\overline{15})$ is negligible comparing  to $H(6)$. 
However, our results are somewhat different from those in Ref.~\cite{Sharma:1996sc}. 

There are some common features between our results and those in Refs.~\cite{Sharma:1996sc,Heav,pole}. The most important one
 is that the vanishing amplitudes  in the Cabibbo allowed decays of $\Xi_c^+ \to \Sigma'^+ \bar{K}^0$ and $\Xi_c^+ \to \Xi'^0 \pi^+$.
 It is clear that  
  the current  experimental data of ${\cal B}(\Xi_c^+ \to \Sigma'^+ \bar{K})/{\cal B}(\Xi_c^+\to \Xi^- 2\pi^+)=(1.0\pm0.5)$
  and ${\cal B}(\Xi_c^+ \to \Xi'^0 \pi^+)/{\cal B}(\Xi_c^+\to \Xi^- 2\pi^+)<0.1$~\cite{pdg} are insufficient to rule out this feature yet.
It is interesting to note that the decay branching ratios given in the various theoretical calculations may not obey the flavor symmetry of
$SU(3)_F$ in general, but they all preserve the isospin symmetry. 
In particular, the isospin relations in the Cabibbo allowed decays can be summarized as follows:
\begin{eqnarray}\label{isoRe}
{\cal B}(\Lambda_{c}^+\to\Delta^{++} K^-) =3{\cal B}(\Lambda_{c}^+\to\Delta^{+} \bar{K}^0)\,,~~~{\cal B}(\Lambda_{c}^+\to\Sigma^{\prime +} \pi^0)={\cal B}(\Lambda_{c}^+\to\Sigma^{\prime 0} \pi^+)\nonumber\\
{\cal B}(\Xi_{c}^0\to\Sigma^{\prime +} K^-)=2 {\cal B}(\Xi_{c}^0\to\Sigma^{\prime 0} \bar{K}^0)       \,,~~~ 
{\cal B}(\Xi_{c}^0\to\Xi^{\prime 0} \pi^0)=\frac{1}{2}{\cal B}(\Xi_{c}^0\to\Xi^{\prime -} \pi^+)\,.
\end{eqnarray}
Similar relations in the singly and doubly-Cabibbo suppressed decays are also expected.

Finally, we explore the decay processes of $\Xi_c^0 \to \Sigma^{\prime 0}K_S/K_L$, which contain  both Cabibbo allowed and doubly-suppressed 
contributions as shown in Table~\ref{table4}, resulting in an asymmetry  due to the interference between the two contributions.
Explicitly, the $K_S-K_L$ asymmetry is found to be
\begin{equation}\label{kaonrela}
{\bf R}\equiv\frac{\Gamma(\Xi_c^0\to \Sigma^{\prime 0} K^0_{S})-\Gamma(\Xi_c^0\to \Sigma^{\prime 0} K^0_{L})}{\Gamma(\Xi_c^0\to \Sigma^{\prime 0} K^0_{S})+\Gamma(\Xi_c^0\to \Sigma^{\prime 0} K^0_{L})}=\frac{(1-s_c^2)^2-(1+s_c^2)^2}{(1-s_c^2)^2+(1+s_c^2)^2}=-0.106\,,
\end{equation}
which is independent of the fitting.
As a consequence,
 the asymmetry in Eq.~(\ref{kaonrela}) provides a clean prediction in the $SU(3)_F$ approach for the charmed baryon decays, which can be tested by
  the experiments in BELLE and BESIII.

\begin{table}
	\begin{center}
		\centering
		\caption{
		Results for $\Xi_c^0 \to \Sigma^{\prime 0} K_S/K_L$ with  $SU(3)_F$. }
		\label{table4}
		\begin{tabular}[t]{ccccc}
			\hline
			channel&$f_{{\bf B}_c{\bf B}_DM} $&$\alpha_{pm}$&$10^3{\cal B}_{pm}$&$10^3{\cal B}_{em}$\\
			\hline
$ \Xi_{c}^{0}  \to  \Sigma'^{0} K_S $&$  - \frac{\sqrt{3} }{3} +\frac{\sqrt{3}  }{3}s_{c}^{2}$&$ -0.88 ^{+ 0.45 }_{- 0.12 }$&$ 0.70 \pm 0.11 $&$ 0.52 \pm 0.04 $\\
$ \Xi_{c}^{0}  \to  \Sigma'^{0} K_L $&$ \frac{\sqrt{3}}{3}+\frac{\sqrt{3} }{3} s_{c}^{2}  $&$ -0.88 ^{+ 0.45 }_{- 0.12 }$&$ 0.87 \pm 0.14 $&$ 0.64 \pm 0.05 $\\

\hline
		\end{tabular}
	\end{center}
\end{table}

\section{Conclusions}
We have studied the decay branching ratios and up-down asymmetries in the charmed baryon  weak decays of
 ${\bf B}_c\to {\bf B}_DM$  based on the flavor symmetry of  $SU(3)_F$. 
 It is interesting to emphasize that these ${\bf B}_c$ decays with the decuplet spin-3/2 baryon receive only non-factorizable contributions. 
 We have shown that our fitting results for ${\cal B}({\bf B}_c\to {\bf B}_DM) $ are consistent with the current experimental data in both
 $pm$ and $em$ schemes. 
In particular,  the $em$ scheme leads to a much smaller number for the $\chi^2$ fit than the $pm$ one, 
resulting in that the predicted values of ${\cal B}({\bf B}_c\to {\bf B}_DM) $ in the 
$em$ scheme contain much less uncertainties than those in the $pm$ one.
To reduce the large uncertainties in the $pm$ scheme, the $SU(3)_F$ breaking effect should be included in the 
amplitude as well when more precision measurements 
of ${\cal B}({\bf B}_c\to {\bf B}_DM) $ are available.
We have  demonstrated that the isospin relations for the decay branching ratios in Eq.~(\ref{isoRe}) are
scheme- and model-independent. 
It is also interesting to note  that the vanishing rates for the 
Cabibbo allowed decays of $\Xi_c^+ \to \Sigma'^+ \bar{K}^0$ and $\Xi_c^+ \to \Xi'^0 \pi^+$
have not been supported by the experimental data yet.

 For the up-down asymmetries, we have found that they are sizable, which are different from the prediction of zero due to
 the vanishing D-wave contributions in the literature.
 In particular, we have obtained that  $\alpha ({\bf B}_c\to {\bf B}_DM)=  -1.00^{+ 0.34}_{-0}$ for all decay modes  in the $em$ scheme, 
 while they range from $-1$ to $-0.42$ at $1\sigma$ level in the $pm$ scheme, consistent with the current only available data of
 $\alpha_{ex}(\Lambda_c^+\to\Xi^{\prime 0}K^+)=-1.00\pm0.34$~\cite{Ablikim:2018bir} for the up-down asymmetry.
 To justify the $SU(3)_F$ approach, we have proposed to search for $\alpha(\Lambda_c^+\to \Delta^{++}K^-)$, which is predicted to be
 $-0.86^{+0.44}_{-0.14}$,  in the future experiments, as the the decay has the largest branching rate among ${\bf B}_c\to {\bf B}_DM$.

 In addition, we have examined the processes of $\Xi_c^0 \to \Sigma^{\prime 0}K_S/K_L$, which contain  both Cabibbo allowed and doubly-suppressed 
contributions. We have  predicted the $K_L-K_S$  asymmetry of ${\bf R}(\Xi_c^0\to \Sigma^{\prime 0} K_S/K_L)$ is $-0.106$, which depends on
 neither model/scheme nor the data fitting. Clearly, this asymmetry is  a clean result in the  $SU(3)_F$ approach, which should be
 tested by the experiments.

\section*{ACKNOWLEDGMENTS}
This work was supported in part by National Center for Theoretical Sciences and
MoST (MoST-104-2112-M-007-003-MY3 and MoST-107-2119-M-007-013-MY3).

\end{document}